\documentclass[useAMS,usenatbib]{mn2e}
\usepackage[dvipsone]{graphicx}
\usepackage{amssymb}
\usepackage{aas_macros}

\defcitealias{FaucherGiguere:2009df}{F-GL10}

\title[Millisecond pulsar gamma-ray anisotropy]{Anisotropies in the gamma-ray sky from millisecond pulsars}
\author[Siegal-Gaskins et al.]{Jennifer M.~Siegal-Gaskins$^{1}$\thanks{E-mail:
jsg@mps.ohio-state.edu}, Rebecca Reesman$^{1}$, Vasiliki Pavlidou$^{2}$\thanks{Einstein (GLAST) Fellow}, \newauthor
Stefano Profumo$^{3}$, and Terry P.~Walker$^{1}$\\
$^{1}$Center for Cosmology and Astro-Particle Physics, The Ohio State University, Columbus, OH 43210 USA\\
$^{2}$Astronomy Department, California Institute of Technology, Pasadena, CA 91125 USA\\ 
$^{3}$Department of Physics and Santa Cruz Institute for Particle Physics, University of California, Santa Cruz, CA 95064 USA}

\begin{document}

\date{Accepted  Received ; in original form }

\pagerange{\pageref{firstpage}--\pageref{lastpage}} \pubyear{2010}

\maketitle

\label{firstpage}
\begin{abstract}
Pulsars emerge in the \emph{Fermi} era as a sizable population of gamma-ray sources. Millisecond pulsars (MSPs) constitute an older subpopulation whose sky distribution extends to high Galactic latitudes, and it has been suggested that unresolved members of this class may contribute a significant fraction of the measured large-scale isotropic gamma-ray background (IGRB).  We investigate the possible energy-dependent contribution of unresolved MSPs to the anisotropy of the \emph{Fermi}-measured IGRB\@.  For observationally-motivated MSP population models, 
we show that the preliminary \emph{Fermi} anisotropy measurement
places an interesting constraint on the abundance of MSPs in the Galaxy and the typical MSP flux, about an order of magnitude stronger than constraints on this population
derived from the intensity of the IGRB alone.  We also examine the possibility of a MSP component in the IGRB mimicking a dark matter signal in anisotropy-based searches, and conclude that the energy dependence of an anisotropy signature would distinguish MSPs from all but very light dark matter candidates.
\end{abstract}


\defcitealias{SiegalGaskins:2010nh}{Siegal-Gaskins et al.~2010}
\defcitealias{Vargas:2010en}{Vargas et al.~2010}

\section{Introduction}

In the era of precision gamma-ray astronomy, with data of unprecedented quality from the Fermi Large Area Telescope \citep[\emph{Fermi}-LAT,][]{Atwood:2009ez} and ground-based Atmospheric Cherenkov Telescopes, including H.E.S.S., VERITAS, and MAGIC, long-standing questions about the high-energy universe might soon be successfully addressed. One of these is the detailed nature and origin of the diffuse gamma-ray emission. The gamma-ray sky is dominated at low Galactic latitudes by a bright diffuse Galactic component, stemming dominantly from processes involving cosmic rays such as inelastic hadronic collisions producing neutral pions, and inverse Compton and bremsstrahlung emission from relativistic cosmic-ray electrons and positrons \citep[see, e.g.,][]{Strong:1998fr}. At high latitudes, the diffuse gamma-ray background is customarily attributed to extragalactic gamma-ray emitters, such as blazars \citep[e.g.,][]{Stecker:1996ma}.  Recent \emph{Fermi}-LAT results, however, indicate that resolved blazars only contribute a small fraction of the observed emission (\citealp{Abdo:2009wu}; however, see also \citealp{Abazajian:2010pc}), in contrast to, e.g., the diffuse X-ray background \citep{2005ARA&A..43..827B, 2006ApJ...645...95H, 2007ApJ...661L.117H}.

Since the discovery of periodic gamma-ray emission from pulsars \citep{1971ICRC....1...63B}, the possibility that this source class contributes non-negligibly to the diffuse gamma-ray emission has been considered \citep{1991JApA...12...17B, 1992ApJ...391..659B, 1997ApJ...476..238B}. Some of the brightest emitters in the \emph{Fermi}-LAT gamma-ray sky are in fact associated with pulsating objects, often corresponding to pulsars observed at radio and X-ray frequencies \citep{Abdo:2009mg}. Compared to its predecessor EGRET, \emph{Fermi} is shedding light not only on young, powerful ``ordinary'' pulsars (with typical rotation periods of the order of 0.01-1 sec and ages ranging between $10^3$ and $10^6$ yr) but also on a distinct class of periodic gamma-ray emitters with much shorter pulsating periods (on the order of a few milliseconds), i.e. millisecond pulsars (MSPs). The characteristic age $\tau_c$ of MSPs, extrapolated from their period and period-derivative, indicates that these objects are much older than ordinary pulsars, with $\tau_c\sim10^{10}$ yr \citep{2009Sci...325..848A}. MSPs are thought to be associated with binary systems, the spin-up of the pulsar period being fueled by accretion of mass and angular momentum from the neutron star companion \citep{1994ARA&A..32..591P, 2001LRR.....4....5L}.  While the determination of the age of MSPs is a debated matter given the highly non-trivial nature of their evolutionary history \citep[see, e.g.,][]{Kiziltan:2009rx}, the significantly longer lifetime of these objects compared to that of ordinary pulsars might offset a birthrate that is necessarily lower (given the binary nature of MSPs), and as a result the MSP contribution to the Galactic gamma-ray luminosity may not be small compared to that of ordinary pulsars.

Despite the dramatic increase in the number of detected gamma-ray pulsars in the \emph{Fermi} era, the bulk of the pulsar contribution to the gamma-ray sky very likely originates from a large population of unresolved sources. For instance, \citet[][hereafter \citetalias{FaucherGiguere:2009df}]{FaucherGiguere:2009df}  examined models for the unresolved MSP population and found that in some optimistic models the MSP contribution to the diffuse background could even be dominant at certain energies. In their ``viable model'' the small set of MSPs detected by \emph{Fermi} imply almost 50k unresolved MSPs. The gamma-ray emission from ordinary pulsars is very likely confined to rather low latitudes \citep[see, e.g.,][]{1981ApJ...247..639H, 1991JApA...12...17B}, reflecting the fact that pulsars are born in the Galactic disk, and that ordinary pulsars are relatively young objects. On the other hand, the product of typical pulsar kick velocities and the characteristic age of MSPs implies a length-scale that is much larger than the thickness of the Galactic plane, suggesting that MSPs should have a broad latitudinal distribution.  This is reflected in the observed latitudinal distribution of ordinary versus millisecond pulsars detected by the \emph{Fermi}-LAT \citep[see Fig.~1 in ][]{Abdo:2009ax}. In this respect, MSPs can contribute to the diffuse gamma-ray emission at high latitudes where the Galactic diffuse component is generally thought to be comparable or sub-dominant with respect to an isotropic extragalactic background.

Interestingly, however, measurements of the spectrum of the large-scale isotropic diffuse gamma-ray background (hereafter IGRB) by \emph{Fermi} find that it is consistent with a power law at energies between 250 MeV and 50 GeV \citep{Abdo:2010nz}, while MSP spectra exhibit a strong cut-off feature at typical energies of a few GeV \citep{2009Sci...325..848A}. This implies that either the MSP contribution is subdominant with respect to the primary IGRB component at these energies, or that a complicated combination of several components with peculiar spectral features -- e.g., a star-forming galaxy component with a feature at $\sim$300 MeV \citep[e.g.,][]{2010ApJ...722L.199F}, a MSP component with a feature at a few GeV, and a hard blazar component dominating at higher energies -- combine in such a way that they appear as an overall almost featureless power-law -- a contrived scenario, but one that cannot be excluded in principle. In either case, it appears that it will be difficult to detect spectrally the presence of a MSP component in the IGRB, although it remains possible to put conservative constraints on the unresolved MSP gamma-ray emission based on IGRB measurements \citepalias[see, e.g.,][]{FaucherGiguere:2009df}.

A powerful tool to investigate the nature of diffuse emission is to explore the intensity variation of the emission in the sky, e.g., via the calculation of an angular power spectrum of anisotropies. Recent theoretical work has generated predictions for the angular power spectrum of the gamma-ray emission originating from several known and proposed source classes. These include confirmed extragalactic gamma-ray populations such as AGN \citep{Ando2007,miniati_koushiappas_di-matteo_07} and star-forming galaxies \citep{ando_pavlidou_09}, as well as dark matter annihilation and decay in extragalactic structures \citep{Ando:2005xg,miniati_koushiappas_di-matteo_07,Ando2007,cuoco_brandbyge_hannestad_etal_08,taoso_ando_bertone_etal_09,Fornasa:2009qh,ibarra_tran_weniger_09,2010MNRAS.405..593Z,Cuoco:2010jb}.  In addition, since the distribution of dark matter subhalos in our Galaxy is quite radially extended, gamma-ray emission from annihilation and decay in Galactic substructure appears remarkably isotropic on large angular scales, although the clustering of dark matter in subhalos leads to small-scale anisotropies.  Consequently, these structures may provide a substantial contribution to anisotropies in the IGRB \citep{siegal-gaskins_08,Fornasa:2009qh,ibarra_tran_weniger_09,Ando2009}.

The combined use of spectral and anisotropy information in the IGRB (the anisotropy energy spectrum) could conceivably help reveal the presence of even a subdominant component in the diffuse emission \citep{SiegalGaskins:2009ux}. In particular, it has been shown that the anisotropy energy spectrum could be a sensitive probe of the presence of a dark matter component in the IGRB \citep{Hensley:2009gh,Cuoco:2010jb}. This technique is also promising for detecting a subdominant MSP contribution to the IGRB, since the emission from unresolved MSPs is expected to feature much stronger anisotropy than the extragalactic component, due to the fact that MSPs are relatively few and nearby, compared to cosmological populations that may constitute the dominant contributors to the IGRB intensity.  

Additional motivation to study the gamma-ray anisotropy properties of MSPs is provided by the potential interference of MSPs with anisotropy-based searches for dark matter. \emph{Fermi} data \citep{2009Sci...325..848A} indicate that the typical gamma-ray MSP spectrum is, in fact, uncomfortably similar in its overall features to what is expected for the annihilation or decay of certain particle dark matter candidates, especially if the dark matter is light ($m_{\rm DM} \lesssim$ few tens of GeV). Furthermore, although the amplitude of anisotropies from dark matter annihilation 
is uncertain, in some scenarios it is expected to be quite large, and thus it is conceivable that a MSP-induced modulation in the anisotropy energy spectrum of the IGRB could be confused with a similar modulation induced by dark matter.

In this paper, we explore the potential of an angular power spectrum measurement of the IGRB to probe the properties of the Galactic MSP population.  We demonstrate the power of this approach for an example class of MSP population models by deriving constraints on those models from the \emph{Fermi} preliminary anisotropy measurement (\citetalias{SiegalGaskins:2010nh}; see also \citetalias{Vargas:2010en}).  The model prescriptions we adopt to describe the intensity and sky distribution of unresolved MSPs are summarized in \S\ref{sec:model},
and our procedure for generating simulated maps of the MSP gamma-ray emission is outlined in \S\ref{sec:simulations}.  In \S\ref{sec:gammarays} we calculate the intensity spectrum and energy-dependent angular power spectrum of the collective unresolved MSP emission for this class of models and discuss those properties in the context of other relevant source classes, including dark matter.  We compare the predicted anisotropy from MSPs to the preliminary \emph{Fermi} measurement of the angular power spectrum of the IGRB and obtain constraints on the properties of the MSP population in \S\ref{sec:obssig}.  We discuss our findings and conclude in \S\ref{sec:discussion}.

\section{Modeling the MSP population}
\label{sec:model}

The properties of the MSP population that affect the measured anisotropy are the sky distribution of MSPs and their flux distribution.  The former is determined by the spatial distribution of MSPs in the Galaxy, while the latter is determined, for a fixed spatial distribution, by the distribution of MSP luminosities.  In this study we adopt models for the gamma-ray MSP population based on the semi-empirical models of \citetalias{FaucherGiguere:2009df}.  
We emphasize, however, that this work is a technique demonstration, and therefore its goal is to show that MSPs could produce an observable anisotropy signal in \emph{Fermi}-LAT data, and that an anisotropy analysis could be used to constrain the collective properties of the Galactic MSP population; not to perform a detailed study of the consistency of a specific model with the data, nor to identify which of several models is preferred by the data.  With that purpose in mind, we fix the values of the parameters controlling the spatial and luminosity distributions of MSPs to those of ``viable'' model \emph{MSP2\_base} of \citetalias{FaucherGiguere:2009df}, and discuss the expected impact of variations in these parameters on our results in \S\ref{sec:discussion}.  

We take the fiducial number of MSPs in the Galaxy $N_{\rm MSP} = 49$k, as in model \emph{MSP2\_base}.  Since the observables considered in our study (high-latitude intensity and angular power) scale straightforwardly with $N_{\rm MSP}$ and the typical flux of an individual high-latitude MSP, $F_{1}$, we also consider the dependence of our results on these parameters in \S\ref{sec:obssig}.

Following \citetalias{FaucherGiguere:2009df}, we describe the MSP spatial distribution with a Gaussian function of radius for the surface density projected on the Galactic plane,
\begin{equation}
\label{eq:radial}
\rho(r) \propto \exp(-r^{2}/2\sigma_{r}^{2}) \qquad  0< r < 100\ {\rm kpc},
\end{equation}
where $r$ is the projected distance from the Galactic Centre in the Galactic plane, $\rho(r)$ is the surface density of MSPs, and $\sigma_r$, taken to be $5$ kpc, characterizes the radial extent of the distribution. The latitude distribution of MSPs is assumed to follow a simple exponential form,
\begin{equation} 
\label{eq:zdist}
N(z) \propto \exp{(-|z|/\langle|z|\rangle)} \qquad  0<z<\infty,
\end{equation}
with the scale height $\langle|z|\rangle=1$ kpc.  

Early work on gamma-ray pulsars \citep[e.g.,][]{1996A&AS..120C..49A} identified the simple empirical relation $L_{\gamma} \propto \sqrt{\dot{E}}$ between the pulsar's gamma-ray luminosity $L_{\gamma}$ and the rate it loses rotational kinetic energy $\dot E=4 \pi^{2} I_{\star} \dot P /P^{3}$, where $P$ and $\dot P$ are the period and time derivative of the period, respectively, and $I_{\star}$ is the moment of inertia of the star.  
However, recent work \citepalias[see, e.g.,][]{FaucherGiguere:2009df} has found that the luminosities of gamma-ray MSPs appear to obey the relation $L_{\gamma} \propto \dot{E}$.
As in  \citetalias{FaucherGiguere:2009df}, we define the MSP gamma-ray energy luminosity (energy per unit time)
\begin{equation}
\label{eq:lum}
L_{\gamma} \equiv \min\{C \dot P^{1/2} P^{-3/2} ,f_{\gamma}^{\rm max} \dot E \},
\end{equation}
where the proportionality constant $C=10^{40.9}$erg $s^{1/2}$ and $f_{\gamma}^{\rm max} = 0.05$ is the assumed maximum fraction of rotational power loss converted into gamma rays.  The integrated photon luminosity (photons per unit time) $L_{\gamma}^{\rm ph}$ above 100 MeV is obtained by assuming an energy spectrum with approximately equal power per decade of energy up to a cutoff energy of $E_{\rm max} \simeq 3$ GeV.  For the model adopted in this work, it is notable that Eq.~\ref{eq:lum} results in the vast majority of MSPs being assigned luminosities according to the $L_{\gamma} \propto \dot{E}$ relation.

For the MSP population, a power-law distribution for the rotation period $P$ is assumed,
\begin{equation}
\label{eq:period}
N(P) \propto P^{-2} \qquad  1.5\ {\rm ms} < P < 60000\ {\rm ms}, 
\end{equation}
and the magnetic field strength $B$ is taken to follow a log normal distribution,
\begin{equation}
\label{eq:bfield}
N(\log B)\propto \exp{(-(\log B - \langle \log B \rangle)^{2}/2\sigma_{\log B}^{2})},
\end{equation}
with $\langle \log B \rangle = 8$ and $\sigma_{\log B}=0.2$ with $B$ in Gauss.  The spin-down rate $\dot{P}$ is determined via the relation $B=3.2 \times 10^{19} (P \dot{P})^{1/2}$ G.

Departing from the \citetalias{FaucherGiguere:2009df} prescription, 
we adopt an empirical prescription for the energy spectra of the MSPs based on the spectra of eight MSPs detected by \emph{Fermi}, reported in \citet{2009Sci...325..848A}.  The differential energy spectra of the \emph{Fermi}-detected MSPs are well-described by a power law truncated by an exponential cutoff,
\begin{equation}
\frac{{\rm d}N}{{\rm d}E} \propto E^{-\Gamma}e^{-E/E_{\rm cut}},
\end{equation}
where $\Gamma$ is the spectral index and $E_{\rm cut}$ is the cutoff energy.  We assume that each spectral parameter, $\Gamma$ and $E_{\rm cut}$, is normally distributed in the MSP population, with mean $\langle \Gamma \rangle$ and $\langle E_{\rm cut} \rangle$ and standard deviation $\sigma_\Gamma$ and $\sigma_{E_{\rm cut}}$, respectively. We use the spectral parameters of the detected MSPs to identify the maximum-likelihood values of these distribution parameters, taking into account the measurement uncertainties for each pulsar. For this procedure we follow the methodology described in \citet{Venters:2007rn}, and obtain the maximum likelihood parameters  $[\langle\Gamma\rangle,\sigma_{\Gamma}] = [1.5, 0.20]$ and $[\langle E_{\rm cut}\rangle,\sigma_{E_{\rm cut}}] = [1.9$ GeV, $0.54$ GeV$]$.  It is notable that the distributions are relatively narrow; in particular they imply that the vast majority of MSPs have cutoff energies between $\sim 1$ and 3 GeV.

We consider two models for the energy spectra of the MSP population.  First we examine the simple case in which every MSP is assumed to have the same energy spectrum, which we denote the \emph{Reference Model}. In this scenario we assign each MSP the maximum-likelihood average spectral parameters, $\Gamma = 1.5$ and $E_{\rm cut} = 1.9$ GeV.  We also examine the impact on our results of allowing the spectral parameters to vary within the MSP population according to the distributions above, and refer to this as the \emph{Spectral Variation Model}.

\section{Simulations}
\label{sec:simulations}

Monte Carlo realizations of the Galactic MSP population were generated by creating mock MSP catalogs with individual MSP parameters drawn from the distributions given in \S\ref{sec:model}.
To assess the statistical variation between realizations, ten Monte Carlo realizations were generated for each case considered.  The HEALPix package \citep{Gorski:2004by} was used to generate maps of the all-sky gamma-ray intensity from unresolved MSPs for each mock catalog.  Maps were constructed at HEALPix order 7 resolution which corresponds to a pixel size of $\sim 0.45^{\circ}$ on a side.  Each MSP was taken to be a point source with no angular extent, and so its flux was assigned to a single pixel.

As we are interested in the emission from unresolved MSPs, we excluded from the sky maps the emission from MSPs in our mock catalogs which would likely have been detected by \emph{Fermi}.  For this purpose we assumed a flux sensitivity of $10^{-8}$ ph cm$^{-2}$ s$^{-1}$ ($E > 100$ MeV; see, e.g., \citealp{Abdo:2009ax}) and excluded individual MSPs exceeding this flux threshold.   
We note that assuming a uniform flux sensitivity for MSPs across the sky is a rough approximation, since point source sensitivity varies with angular position due to exposure and foreground contamination, and also depends on the individual source spectrum.  Although we emphasize that this approximation is inadequate for assessing individual source detectability or completeness, it is sufficient for the purpose of removing bright sources which are likely to be resolved and would otherwise bias our prediction for the statistical properties of the diffuse emission.  Choosing to exclude these sources is conservative, since including bright MSPs would lead to a larger predicted intensity and a larger contribution to the anisotropy of the IGRB from MSPs, and as a result to stronger constraints on MSP population models.  With this criterion we find that $\sim10$ of every $10$k MSPs in our Reference Model are detectable over the entire sky. Note that the parameter $N_{\rm MSP}$ corresponds to the total number of MSPs in the Galaxy and therefore includes the detectable sources, although our analysis is performed on maps of the unresolved sources only.

\section{Gamma-ray emission from the unresolved MSP population}
\label{sec:gammarays}

\begin{figure*}
\includegraphics[width=0.65\textwidth]{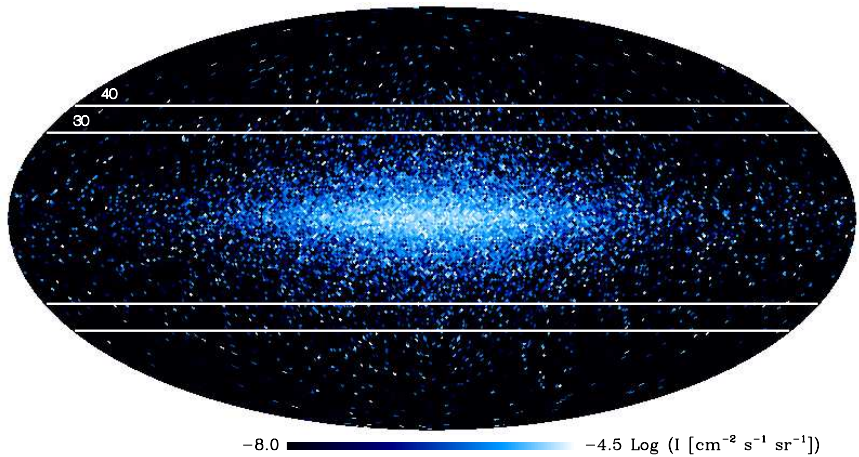}
\caption{MSP gamma-ray intensity integrated from 0.1 to 10 GeV for one realization of the Reference Model.  The map is shown in Galactic coordinates with the boundaries of the latitude masks excluding $|b| < 30^\circ$ and $|b| < 40^\circ$ marked.  For this figure the map resolution was degraded to improve the visibility of MSPs and illustrate their sky distribution; however, all calculations were performed on the high-resolution maps as described in the text.\label{fig:intensmaps}}\vspace{\baselineskip}
\end{figure*}
 
\subsection{Sky distribution of gamma rays from MSPs}

We examined the constraints obtainable on the MSP population from the intensity and anisotropy properties of the \emph{Fermi-}measured IGRB\@, so we selected high-latitude sky regions by excluding Galactic latitudes $|b| < 30^\circ$.  This choice matches the latitude mask applied in the \emph{Fermi} angular power spectrum analysis \citepalias{SiegalGaskins:2010nh}.  The latitude dependence of the results was studied by comparing the results using a mask excluding $|b| < 40^\circ$.  The choice to apply a very generous mask to the Galactic plane also enables comparison of the high-latitude contribution of MSPs to the \emph{Fermi-}measured IGRB intensity.

An all-sky map of the gamma-ray intensity for one realization of the MSP population Reference Model defined in~\S\ref{sec:model} with $N_{\rm MSP} = 49$k is shown in Figure~\ref{fig:intensmaps}.  Emission from individual MSPs with fluxes above the detectability threshold is not shown.  
MSPs outside of the latitude mask boundaries (marked by lines) are evident, implying a MSP contribution to high-latitude diffuse emission.  

\subsection{Intensity energy spectra}

\begin{figure}
\includegraphics[width=0.47\textwidth]{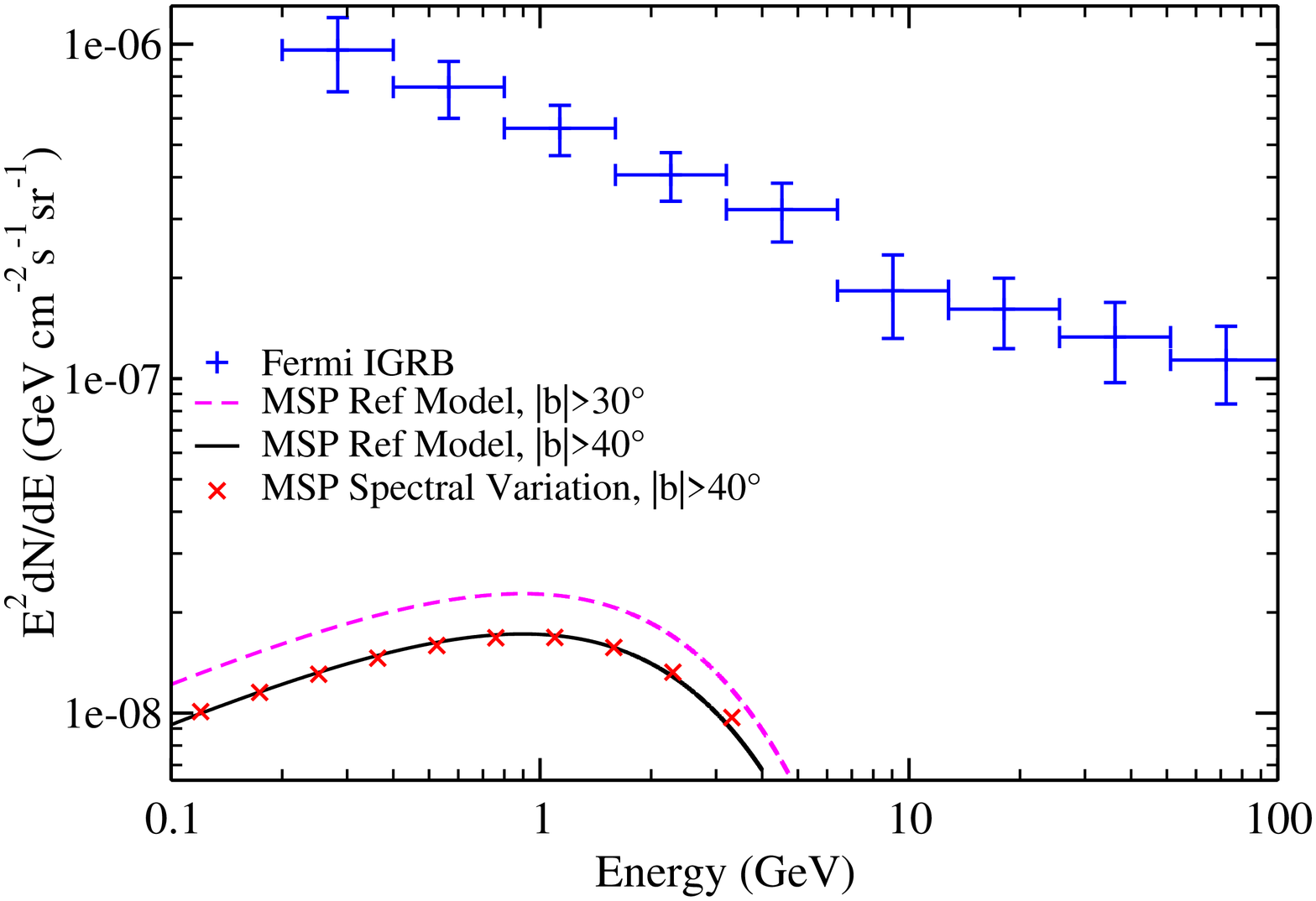}
\caption{Average intensity energy spectra of the MSP Reference Model (solid black line) and Spectral Variation Model (red x's) for $|b|>40^{\circ}$. The intensity spectrum of the Spectral Variation Model differs negligibly from that of the Reference Model.  The average intensity of the MSP Reference Model for $|b|>30^{\circ}$ (dashed magenta line) is also shown.   The collective high-latitude intensity of the MSPs is more than an order of magnitude smaller than the \emph{Fermi}-measured IGRB intensity (blue crosses) at all energies.\label{fig:intensity}}
\end{figure}

The intensity energy spectrum of the gamma-ray emission from MSPs outside each latitude mask for the Reference Model is compared with the \emph{Fermi-}measured IGRB intensity spectrum \citep{Abdo:2010nz} in Fig.~\ref{fig:intensity}.  The normalization of the MSP intensity outside each mask was obtained by averaging over 10 realizations, and the spectral parameters of each MSP in the Reference Model were fixed to the maximum likelihood values.  The average intensity of the emission from unmasked MSPs is a factor of $\sim 2$ larger when excluding only $|b| < 30^\circ$ than when excluding $|b| < 40^\circ$, but in both cases is more than an order of magnitude smaller than the IGRB at all energies.  This model is consistent with the measured IGRB, but the overall intensity does not provide a meaningful constraint on the population.

In the Reference Model we adopted the simplifying assumption that all MSPs share the same energy spectrum.  To test the validity of this assumption, in Fig.~\ref{fig:intensity} we compare the collective intensity spectrum of the MSP population, averaged over $|b| > 40^\circ$, for the Spectral Variation Model and the Reference Model.  The differential intensity ${\rm d}N/{\rm d}E$ of each realization of the Spectral Variation Model was obtained by generating a map for each logarithmic energy bin containing the integrated intensity of each MSP, given its spectral parameters, and then dividing the integrated intensity by the energy bin size $\Delta E$.  The points shown represent the average intensity outside the mask in each energy bin of 10 Monte Carlo realizations of the Spectral Variation Model.  To good approximation, the collective intensity energy spectrum of the Spectral Variation Model matches that of the Reference Model, with a small deviation from the Reference Model spectrum evident only at the highest energy bin ($E \sim 3$ GeV).

\begin{figure}
\includegraphics[width=0.47\textwidth]{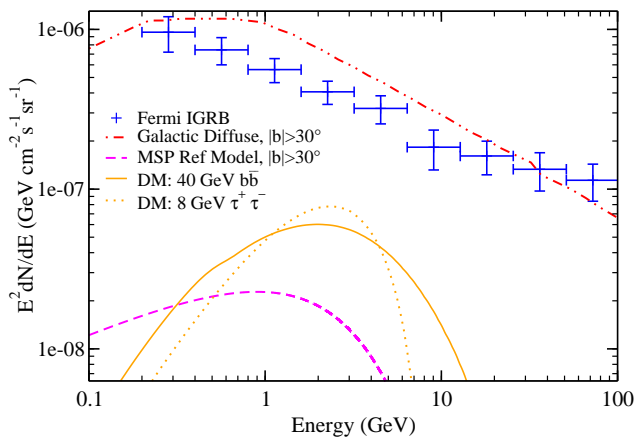}
\caption{Average intensity spectra of the MSP Reference Model for $|b| > 30^{\circ}$ (dashed magenta line) compared with the IGRB intensity (blue crosses), the Galactic diffuse emission for $|b| > 30^{\circ}$ \citep[dot-dashed red line, from][]{Cuoco:2010jb}, and two benchmark dark matter models.  The two dark matter models correspond to an 8 GeV particle pair-annihilating preferentially into $\tau^+\tau^-$ at a rate $\langle\sigma v\rangle=1\times10^{-26}{\rm cm}^3{\rm s}^{-1}$ (dotted yellow line), and to a 40 GeV particle annihilating into $b\bar b$ with $\langle\sigma v\rangle=3\times10^{-26}{\rm cm}^3{\rm s}^{-1}$ (solid yellow line).\label{fig:intensity2}}
\end{figure}

In Fig.~\ref{fig:intensity2} we compare the MSP intensity for $|b|>30^{\circ}$ to the Galactic diffuse emission for $|b|>30^{\circ}$ \citep[from the model used in][]{Cuoco:2010jb}.  At these latitudes, the intensity of the Galactic diffuse emission from cosmic-ray interactions with the interstellar gas and photon fields is comparable to that of the IGRB, and the MSP emission is subdominant with respect to both of these signals.  However, the Galactic diffuse emission is not expected to contribute significantly to the anisotropy on angular scales of $\lesssim 1-2^{\circ}$, corresponding to multipoles $\ell \gtrsim 100$ \citep[see, e.g.,][]{Cuoco:2010jb}, and therefore MSPs could be a dominant contributor to the anisotropy of the high-latitude diffuse emission while remaining a subdominant contributor to the intensity.  

Fig.~\ref{fig:intensity2} also compares the intensity spectrum of MSPs to that of the high-latitude emission predicted for two example dark matter models, chosen because their energy spectra bear some resemblance to the collective MSP energy spectrum.  We do not resort to any specific particle physics setup in the choice of the models. Rather, we specify a dominant pair-annihilation final state, the particle mass, and the rate of pair-annihilation. One of the dark matter models corresponds to a dark matter particle with a mass of 8 GeV and a cross section $\langle\sigma v\rangle=1\times10^{-26}{\rm cm}^3{\rm s}^{-1}$, for which the dominant annihilation final state is a pair of $\tau$ leptons.  This model was chosen to align with that found in the analysis of ~\citet{Hooper:2010mq} to best fit a gamma-ray excess claimed to exist in the innermost 2 degrees in the direction of the Galactic Centre (see also \citealt{Abazajian:2010zy} for an interpretation of that signal as MSP emission). We also compare a second dark matter model, with a mass of 40 GeV and a pair-annihilation cross section $\langle\sigma v\rangle=3\times10^{-26}{\rm cm}^3{\rm s}^{-1}$, for which the dominant annihilation final state is bottom quarks. This second model can be regarded as a prototypical light bino-like dark matter candidate from the minimal supersymmetric extension of the Standard Model, with a cross section that would allow for thermal production of the correct universal dark matter density.  The intensity of the dark matter emission for these two models corresponds to that predicted for the high-latitude signal from annihilation in Galactic dark matter subhalos in model A1 of \citet{Ando2009}, assuming the particle properties for each model specified above.  

Dark matter annihilation or decay in Galactic substructure may generate a significant level of anisotropy in the IGRB with an energy dependence similar to that from MSPs due to their similar energy spectra.  Although the detailed shapes of the energy spectra of the dark matter models shown in Fig.~\ref{fig:intensity2} differ from that of the collective MSP emission, the energy range at which both of these possible contributors become most prominent in the IGRB, as well as their cutoff energies, are similar.  Since an anisotropy analysis requires large photon statistics to robustly measure small anisotropies, the number of energy bins in which a measurement can be made with \emph{Fermi}-LAT is limited, and therefore it may be difficult to localize features in the energy dependence of the anisotropy.  Consequently, there remains the possibility that a MSP-induced anisotropy in the IGRB could be confused with a similar signal from dark matter annihilation.  However, we stress that only a signal from very light dark matter candidates is likely to exhibit a spectral cutoff at sufficiently low energies to effectively mimic MSPs in an anisotropy measurement.

\subsection{Angular power spectra}

We consider the angular power spectrum of intensity fluctuations $\delta I (\psi) = (I(\psi) - \langle I \rangle)/\langle I \rangle$ where $I(\psi)$ is the intensity in the direction $\psi$, and $\langle I \rangle$ is the average intensity over the unmasked region of the sky.  The angular power spectrum is calculated by expanding $\delta I$ in spherical harmonics $\delta I =  \sum_{\ell,m} a_{\ell,m} Y_{\ell,m} (\psi)$ to obtain the coefficients $C_{\ell} = \langle | a_{\ell,m}|^{2} \rangle$.  Since a fluctuation map is dimensionless, its angular power spectrum characterizes the angular distribution of the emission, independent of its overall intensity.  

\begin{figure}
\includegraphics[width=0.47\textwidth]{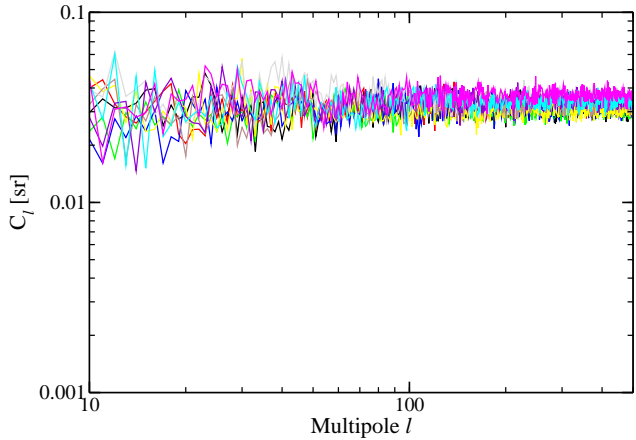}
\caption{Angular power spectrum of the Reference Model, with Galactic latitudes $|b|<40^{\circ}$ masked.  Each line corresponds to one of ten realizations; the variation between realizations is small.  The $C_{\ell}$ are remarkably constant in multipole, which is consistent with the angular power spectrum of an uncorrelated distribution of point sources.\label{fig:refaps}}
\end{figure}

We calculate the angular power spectrum of the emission from MSPs from the simulated sky maps using HEALPix.  The angular power spectra are calculated on the cut sky, after removing the monopole and dipole components.  To approximately correct for the power suppression due to masking, the angular power spectra of the cut sky are divided by the fraction of the sky outside the mask, $f_{\rm sky}$.  This approximation is valid at multipoles $\ell \gtrsim 100$.

The angular power spectra of 10 Monte Carlo realizations of the Reference Model are shown in Fig.~\ref{fig:refaps}, calculated with a mask excluding $|b| < 40^{\circ}$.  The scatter between realizations is small, with each realization generating an angular power spectrum $C_{\ell}$ approximately constant in multipole with a value $0.03 \lesssim C_{\ell} \lesssim 0.04$ for $\ell \gtrsim 100$.  The multipole-independence of $C_{\ell}$ is characteristic of the power spectrum of Poisson noise (shot noise) $C_{\rm P}$, which arises from an uncorrelated distribution of sources.  Noting that the angular power spectrum from MSPs at high latitudes appears to be dominated by the Poisson contribution, we hereafter make the approximation that the angular power from MSPs is constant in multipole, and identify $C_{\rm P}$ as the average of $C_{\ell}$ over $\ell=50$ to $\ell=150$.

\begin{figure}
\includegraphics[width=0.47\textwidth]{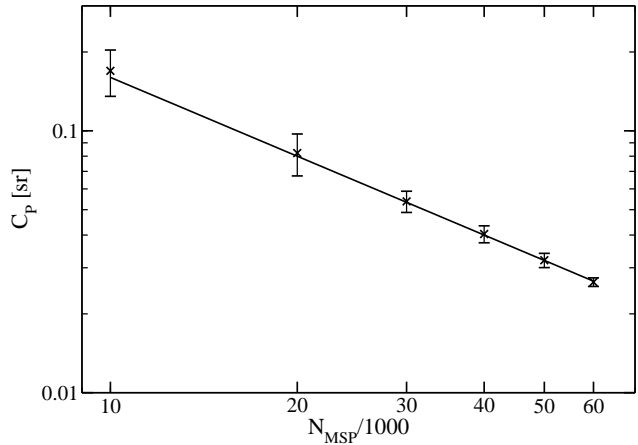} 
\caption{Dependence of Poisson angular power $C_{\rm P}$ on $N_{\rm MSP}$ for the Reference Model, with Galactic latitudes $|b| <40^{\circ}$ masked.  The error bars represent the standard deviation of the $C_{\rm P}$ from ten realizations, and the solid line illustrates the expected relation $C_{\rm P} \propto 1/N_{\rm MSP}$.\label{fig:ClvsNMSP}}
\end{figure}

The Poisson contribution to the power spectrum scales inversely to the number density of sources, i.e., $C_{\rm P} \propto 1/\mathcal{N}$, where $\mathcal{N}$ is the number of sources per solid angle.
Figure~\ref{fig:ClvsNMSP} illustrates the dependence of the angular power on the total number of MSPs in the model $N_{\rm MSP}$.  Here we adopted the Reference Model but varied $N_{\rm MSP}$; for each value of $N_{\rm MSP}$, maps for ten realizations of the MSP emission were generated.  As before, MSPs with individual fluxes above the detection threshold were not included in the maps.  The average $C_{\rm P}$ of the ten maps for each $N_{\rm MSP}$ is shown in the figure.  As expected for a Poisson-like source distribution, the angular power scales inversely with the number of MSPs.

It is important to confirm that the expected variation of the spectral parameters of MSPs within the Galactic population, in particular the distribution of cutoff energies, does not introduce an energy dependence into the angular power spectrum.  For a source distribution that is independent of energy (i.e., a source class in which each member has the same observed intensity energy spectrum), the fluctuation angular power spectrum is also energy-independent.  Energy dependence of anisotropy indicates a change in the spatial distribution of the contributing sources with energy, and can be used to identify the presence of multiple populations or populations whose properties vary significantly with energy \citep{SiegalGaskins:2009ux}.  The angular power spectra of some astrophysical gamma-ray source populations are expected to exhibit a mild, characteristic energy dependence due to, e.g., large variations in the spectral properties of individual members of a source class, attenuation by interactions with the extragalactic background light (EBL), and redshifting \citep[see, e.g.,][]{Ando:2005xg,Zhang:2004tj}.  In contrast, the angular power spectrum of emission from Galactic dark matter annihilation or decay would be constant in energy since the energy spectrum is fixed for a given dark matter particle, and emission from Galactic sources is not subject to redshifting and EBL attenuation.

\begin{figure}
\includegraphics[width=0.47\textwidth]{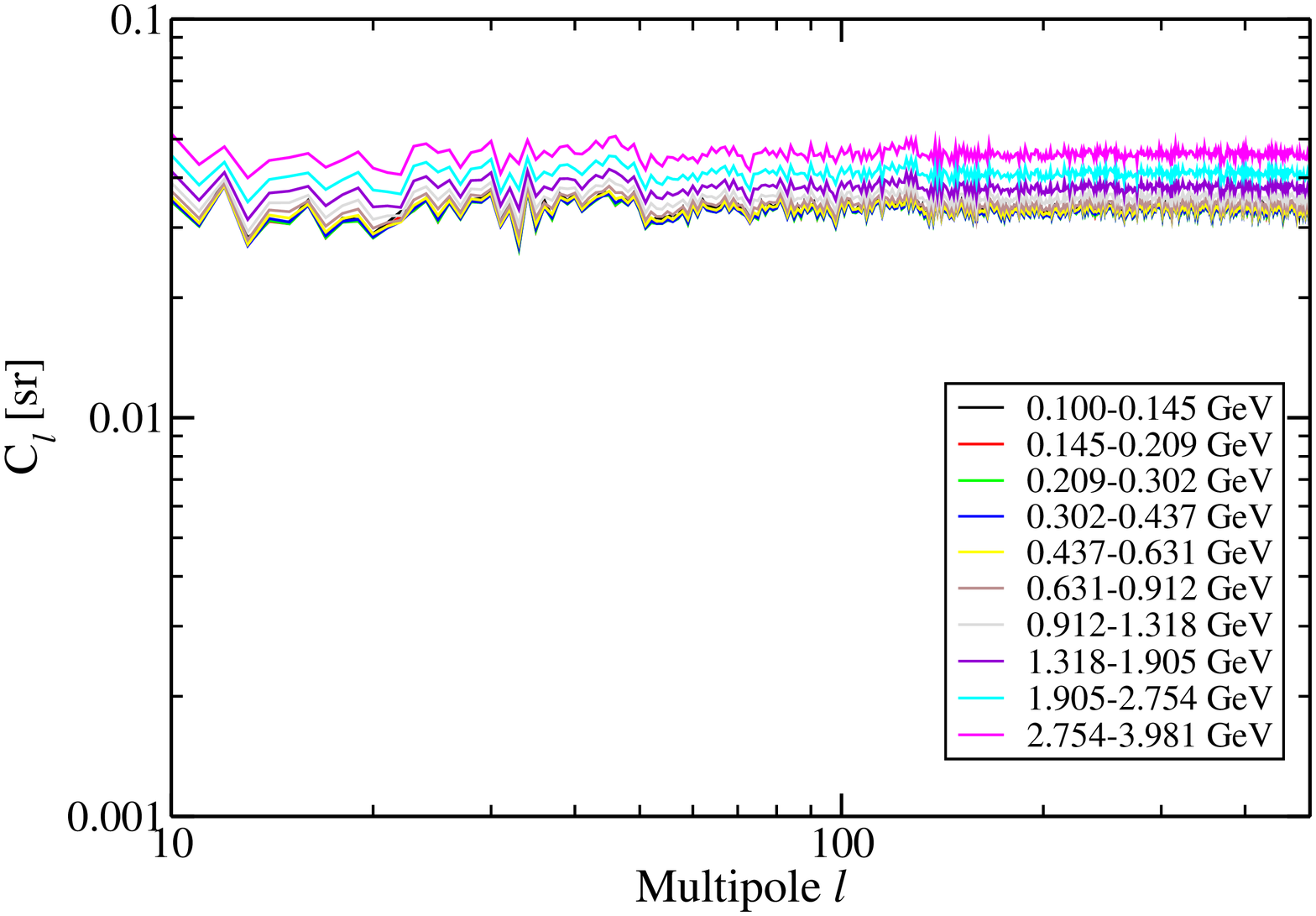}
\caption{Energy dependence of the angular power spectrum of the emission from the MSP Spectral Variation Model.  The angular power spectrum shown for each energy bin has been averaged over ten Monte Carlo realizations, and was calculated with $|b|<40^{\circ}$ masked.\label{fig:vary_aps}}
\end{figure}

The energy dependence of the angular power spectrum of the Spectral Variation Model is examined in Fig.~\ref{fig:vary_aps}.  The angular power spectrum of the MSP emission was calculated in each of 10 logarithmically-spaced energy bins, and then averaged over ten realizations.  The amplitude of the angular power spectrum at a given multipole varies by less than a factor of 2 over the energy range considered, and the variation is noticeable only for the highest energy bins ($E \gtrsim 1$ GeV).  The slight increase in the angular power at high energies is expected for the Spectral Variation Model since, due to the variation in cutoff energies within the MSP population for this scenario, some MSPs no longer contribute to the intensity at the highest energy bins, decreasing the number density of sources $\mathcal{N}$ and thereby increasing the angular power $C_{\rm P}$.

\section{Observational constraints}
\label{sec:obssig}

We now illustrate the potential of anisotropy measurements to constrain the properties of the MSP population by comparing the predictions of our population model to the preliminary \emph{Fermi} anisotropy measurement, and deriving constraints on the abundance and emission properties of gamma-ray MSPs in the Galaxy.  We impose the requirement that the MSPs do not overproduce the IGRB intensity or anisotropy in the energy range from 1 to 2 GeV, and determine the parameter space of MSP models which is compatible with this constraint.  Since confirmed source populations other than MSPs (e.g., blazars) are expected to contribute both intensity and anisotropy to the IGRB, allowing MSPs to contribute all of the measured intensity or anisotropy is a conservative choice.

To assess the contribution of an individual source class to the total measured IGRB anisotropy $C_{\ell, {\rm tot}}$ we construct the dimensionful angular power spectrum of the intensity by multiplying the fluctuation angular power spectrum of a single source class $C_{\ell}$ by the mean intensity $\langle I \rangle$ of that source class squared, $\langle I \rangle^{2} C_{\ell}$, where the mean intensity is calculated on the unmasked region of the sky.  

Preliminary results from \emph{Fermi} indicate that the IGRB angular power spectrum is approximately constant for $\ell \gtrsim 100$, so we identify that measurement as $C_{\rm P,IGRB}$.  The preliminary \emph{Fermi} measurement of the IGRB angular power spectrum for the energy range of 1-2 GeV, weighted by the mean intensity squared, is $(I^{2}\, C_{\rm P})_{\rm IGRB} \simeq 6.2 \times 10^{-18}$ (cm$^{-2}$ s$^{-1}$ sr$^{-1}$)$^{2}$ sr \citepalias{SiegalGaskins:2010nh}.  The value of $(I^{2}\, C_{\rm P})_{\rm IGRB}$ used here represents the mean of the data points from multipoles of $\ell=100$ to $\ell=200$.  We constrain the MSP contribution to the anisotropy at the 2-$\sigma$ level, i.e., $I^{2}_{\rm tot, MSP}\,C_{\rm P, MSP} \leq (I^{2}\, C_{\rm P})_{\rm IGRB} + 2\sigma_{\rm aniso}$ for 1-2 GeV, where $\sigma_{\rm aniso}$ denotes the mean reported uncertainty on the data points.

Similarly, we derive a constraint from the intensity of the IGRB by requiring $I_{\rm tot, MSP} \leq I_{\rm IGRB} + 2\sigma_{\rm I}$, integrated from 1 to 2 GeV.  \emph{Fermi's} measurement of the IGRB energy spectrum is consistent with a power-law with spectral index $\Gamma_{\rm IGRB} = 2.41$ and $I(>100 {\rm MeV}) = 1.03 \times 10^{-5}$ cm$^{-2}$ s$^{-1}$ sr$^{-1}$  \citep{Abdo:2010nz}\footnote{We note that the intensity of the IGRB used in the anisotropy measurement is not equivalent to $I_{\rm IGRB}$.  The value of $I_{\rm IGRB}$ was determined by a fitting procedure to remove any spatially-dependent components, while the mean intensity of the emission for the anisotropy measurement was calculated by simply applying a mask excluding $|b|<30^{\circ}$ and masking point sources.  The procedure we used to calculate the mean intensity of the MSPs, i.e., excluding $|b|<30^{\circ}$ and emission from individual MSPs above the threshold flux, thus closely corresponds to the approach used in the \emph{Fermi} anisotropy analysis.}, so we adopt this parameterization to determine the integrated intensity of the IGRB from 1 to 2 GeV, $I_{\rm IGRB}$.  The parameter $\sigma_{\rm I}$ is the reported uncertainty in the normalization of the power-law fit to the IGRB intensity.

For the MSP intensity and anisotropy, we adopt the Reference Model and mask $|b| < 30^{\circ}$ to match the latitude mask used in the preliminary \emph{Fermi} anisotropy measurement.  To explore the dependence of the results on the gamma-ray flux distribution of the MSPs, we define $F_{1} \equiv  I_{\rm tot, MSP}\,\Omega_{\rm sky}/(f_{\rm out}\,N_{\rm MSP})$ to parameterize the typical flux contributed by a MSP outside of the mask.  The parameter $I_{\rm tot, MSP}$ is the mean intensity from all MSPs outside the mask, $\Omega_{\rm sky}=4\pi f_{\rm sky}$ is the solid angle of the unmasked sky, and $N_{\rm MSP}$ is the total number of MSPs in the model, as before.  The parameter $f_{\rm out}$ is the average fraction of MSPs outside the mask, which is determined by the spatial distribution adopted for the MSPs.  We calculated $f_{\rm out} = 0.03$ by averaging the fraction of MSPs with $|b| > 30^{\circ}$ in 10 Monte Carlo realizations.  The intensity $I_{\rm tot, MSP}$ and angular power of intensity fluctuations $C_{\rm P, MSP}$ for the MSPs are normalized using the fiducial Reference Model values for $N_{\rm MSP}$ and $F_{1}$.  The intensity of the MSP model then scales as $I_{\rm tot, MSP} \propto F_{1}N_{\rm MSP}$ and the anisotropy as $C_{\rm P, MSP} \propto 1/N_{\rm MSP}$.

\begin{figure*}
\vspace{2.5\baselineskip}
\includegraphics[width=0.65\textwidth]{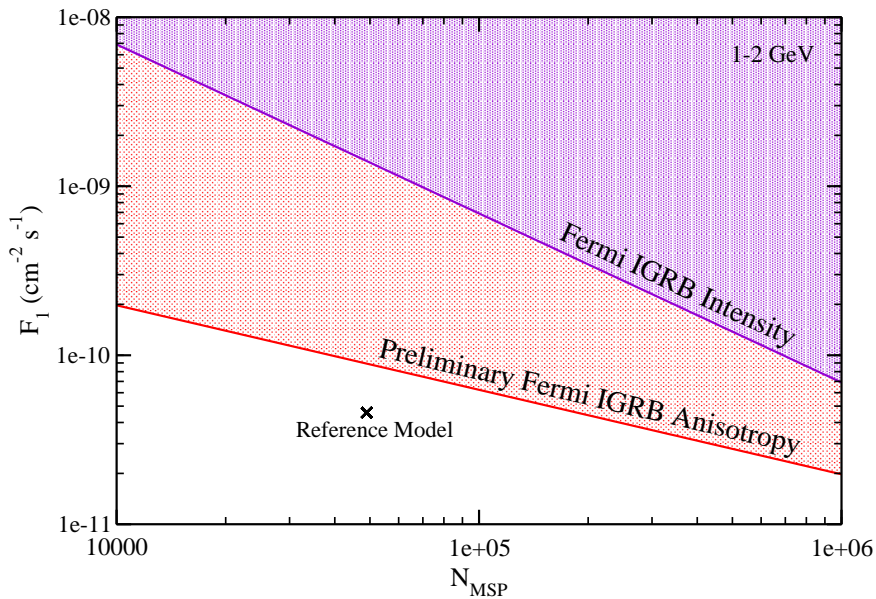}
\caption{Constraints on MSP population properties from the intensity and anisotropy of the IGRB in the energy range 1-2 GeV.  MSP models above the lines exceed the measured value of the total IGRB intensity or anisotropy plus 2$\sigma$.  The reference values for the parameters $F_{1}$ and $N_{\rm MSP}$ are marked.\label{fig:constraints}}
\end{figure*}

Figure~\ref{fig:constraints} shows the regions of the MSP model parameter space excluded by the \emph{Fermi} measurement of the IGRB intensity and preliminary measurement of the IGRB anisotropy.  The fiducial values of the Reference Model are compatible with both the intensity and anisotropy constraints, but an increase in $F_{1}$ of only a factor of two would violate the anisotropy constraint, while the intensity constraint would allow $F_{1}$ of more than an order of magnitude greater, assuming the fiducial $N_{\rm MSP}$.  In general, the anisotropy constraint is significantly stronger than the intensity constraint for MSP fluxes and abundances near the fiducial values.

With regard to the intensity constraint, we caution that our constraint was derived under the assumption that the high-latitude MSP intensity would appear as a contribution to the measured IGRB intensity.  The \emph{Fermi} IGRB measurement \citep{Abdo:2010nz} represents the mean intensity of the all-sky emission that appears to be isotropic on large angular scales.  The high-latitude emission from MSPs as calculated in the present study is not fully isotropic on large angular scales, but rather may exhibit a mild gradient away from the Galactic plane.  Consequently, although unresolved source populations such as MSPs were not explicitly included in the models used in the IGRB intensity spectrum analysis, some or all of the unresolved MSP emission could have been excluded from the reported measurement of the IGRB due to correlations between its large-scale spatial distribution and that of the modeled components.

\section{Discussion}
\label{sec:discussion}

Although tens of thousands of gamma-ray MSPs are expected to live in the Galaxy, only a few tens of MSPs have now been detected individually at gamma-ray energies, presenting a significant challenge to constructing an accurate population model for this source class.  For the purpose of studying the potential contribution of Galactic MSPs to the anisotropy and intensity of the IGRB, we adopted the semi-empirical models outlined in \citetalias{FaucherGiguere:2009df} for the spatial and luminosity distributions of this population.  

The spatial distribution we used is one that is commonly assumed in semi-analytic population studies of MSPs. This spatial model is almost certainly too simplistic, however improved versions require population synthesis models for Galactic compact objects, including a detailed treatment of kinematics in the Galactic gravitational potential and of natal supernova kicks, which generally involve severe uncertainties.  Should additional features in the large-scale spatial distribution of MSPs be robustly predicted by such studies, these could be used as observational signatures tracing a MSP component in the diffuse emission.  In general, large angular scale features can provide key information to help identify the origin of the observed emission, especially for distinguishing a Galactic dark matter signal, which is expected to display a spherical symmetry about the Galactic Centre on large angular scales, and other Galactic source populations, including MSPs, which instead are typically symmetric about the Galactic plane \citep{Hooper:2007be,ibarra_tran_weniger_09,Malyshev:2010zzc}.  

Even without the aid of more detailed models to pin down large-scale features in diffuse emission from MSPs, the general trend of the dependence of the small-scale anisotropy on the spatial distribution of MSPs can be predicted.  We showed that the Poisson contribution to the angular power spectrum $C_{\rm P}$ is the dominant contribution to the total angular power from high-latitude MSPs for multipoles $\ell \gtrsim 100$.  Since $C_{\rm P}$ is inversely proportional to the number density of sources per solid angle, adopting a model that results in a smaller number of MSPs per solid angle in the high-latitude sky regions we considered will lead to a correspondingly larger $C_{\rm P}$.

We demonstrated that, for our Reference Model, the fiducial values for the abundance and average high-latitude MSP flux imply a non-trivial contribution from MSPs to the measured angular power spectrum of the IGRB, despite their contribution to the IGRB intensity of only a few percent.  
In our treatment we focused on the properties of the MSP population which directly impact the IGRB observables.  We defined the parameter $F_{1}$ to describe the mean flux of a MSP with $|b|>30^{\circ}$ in order to encapsulate the information contained in the spatial and luminosity distributions of the MSPs.  Casting our results in terms of $F_{1}$ also provides a means of addressing the uncertainty in the gamma-ray efficiency of MSPs, since for MSPs obeying the relation $L_{\gamma} \propto \dot{E}$, as the overwhelming majority of MSPs in our adopted model do, the gamma-ray flux is linearly proportional to the assumed efficiency factor $f_{\gamma}^{\rm max}$ (see Eq.~\ref{eq:lum}).  
It is important to note that since the luminosity distribution is determined by the distributions of rotation periods and magnetic fields within the population, models which predict alternative distributions for these parameters would likely result in a different predicted $C_{\rm P}$.

To model the energy spectra of the MSP population, we used an empirical approach, determining the spectral shape and expected distribution of spectral parameters from the spectral properties of a small set of \emph{Fermi}-detected MSPs.  Based on that sample, we demonstrated that the dimensionless angular power spectrum of intensity fluctuations from MSPs exhibits only a very mild energy dependence due to spectral variation within the population, and that making the approximation that all MSPs have identical energy spectra had a negligible impact on the predicted anisotropy from this population.  
Although the energy-dependence of the dimensionless anisotropy from MSPs alone is minimal, the \emph{dimensionful} contribution of MSPs to the IGRB anisotropy is strongly energy-dependent, since it depends on the relative contribution of MSPs to the IGRB intensity at each energy.  
To place constraints, we considered the measured IGRB anisotropy at energies from 1 to 2 GeV, which approximately maximizes the MSP contribution to the IGRB intensity for our assumed MSP energy spectrum. 

A potential challenge for interpreting the results of an anisotropy measurement is correctly identifying the source populations contributing to the anisotropy.  Examining the energy dependence of the total anisotropy could help distinguish the contributions of different source populations by taking advantage of differences in their collective energy spectra, but requires that plausible contributing source classes have distinct energy spectra.  In this sense, MSPs could in principle be difficult to distinguish from dark matter annihilation or decay in an anisotropy measurement, due to a general similarity in the shapes of the MSP and dark matter energy spectra for some dark matter models (see Fig.~\ref{fig:intensity2}).  
However, the similarity is only likely to be problematic for very light dark matter candidates with masses $\lesssim$ a few tens of GeV, and therefore MSPs are unlikely to interfere with anisotropy-based searches in gamma rays for the majority of dark matter candidates.  

In spite of a preliminary \emph{Fermi} detection of anisotropies in the IGRB at energies of a few GeV, the sensitivity of anisotropy measurements at higher energies is limited by decreasing photon statistics, and consequently constraints on anisotropies in the IGRB at the level detected in the 1 to 2 GeV band are not yet available for energies much above this range.  We anticipate that future measurements with improved statistics will be better able to constrain the contributions of specific source classes to gamma-ray emission using energy-dependent anisotropy.  Moreover, although we focused on the angular power spectrum of the diffuse emission, we note that a complementary anisotropy statistic, the 1-pt flux PDF, could also be used to help disentangle the contributions of multiple source populations to diffuse emission (\citealt{lee_ando_kamionkowski_09,dodelson_belikov_hooper_etal_09}; \citetalias{FaucherGiguere:2009df}; \citealt{Baxter:2010fr}).  

Although we have not considered these signals here, anisotropy signatures due to interactions of high-energy $e^{\pm}$ injected by MSPs may also be imprinted in emission at lower energies.  This possibility has been pointed out and studied in the context of high-energy charged particles produced in dark matter annihilation in Galactic subhalos, which could lead to signatures in synchrotron emission at radio frequencies from the interactions of $e^{\pm}$ in Galactic magnetic fields \citep{zhang_sigl_08}, and in low-latitude gamma-ray emission from the inverse Compton up-scattering of interstellar radiation field photons by the injected high-energy $e^{\pm}$ \citep{Zhang:2010mg}.

As more individual MSPs are detected at gamma-ray energies, a more accurate population model will emerge.  Anisotropy studies can offer complementary information to that obtained from studies of detected sources by probing the collective properties of the unresolved members of a source class.
For MSPs, which are a source class with relatively few individually detected members, anisotropy analysis could be an important technique to improve our understanding of the characteristics of MSPs at the population level, including their abundance and spatial distribution in the Galaxy. 

\section*{Acknowledgments}
We acknowledge Tim Linden, Kohta Murase, Marco Pierbattista, and Todd Thompson for helpful discussions.  This work was partially supported by NASA through the Fermi GI Program grant number NNX09AT74G.  JSG was supported in part by NSF CAREER Grant  PHY-0547102 (to John Beacom).  RR and TW are supported by DOE grant DE-FG02-91ER406.  JSG, RR, and TW also acknowledge support from The Ohio State University Center for Cosmology and Astro-Particle Physics.  VP acknowledges support for this work provided by NASA through Einstein Postdoctoral Fellowship grant number PF8-90060 awarded by the Chandra X-ray Center, which is operated by the Smithsonian Astrophysical Observatory for NASA under contract NAS8-03060.  SP acknowledges support from the National Science Foundation, award PHY-0757911-001, and from an Outstanding Junior Investigator Award from the Department of Energy, DE-FG02-04ER41286.

\bibliographystyle{mn2e}
\bibliography{references}

\label{lastpage}

\end{document}